\documentclass[conference]{IEEEtran}
\usepackage[a4 paper, top=0.75in, bottom=1in, left=0.625in, right=0.625in]{geometry}
\usepackage{graphicx}
\usepackage{amsfonts}
\usepackage{amsthm}
\usepackage{color}
\usepackage{amsmath}
\usepackage{subfigure}
\usepackage{algorithm}
\usepackage{array}
\usepackage{amssymb}
\usepackage{enumitem}
\usepackage{epstopdf}
\usepackage{dsfont}  
\usepackage{mdwmath}
\usepackage{mdwtab}
\usepackage{mathtools}
\usepackage{multirow}

\ifCLASSINFOpdf
\else
\fi
\usepackage{cite}

\begin{document}
\hyphenation{op-tical net-works semi-conduc-tor}
\renewcommand{\thefootnote}{}
\title{Mobility and Coverage Evaluation of New Radio PDCCH for Point-to-Multipoint Scenario } 
\author{\IEEEauthorblockN{Hongzhi~Chen\IEEEauthorrefmark{1}, De~Mi\IEEEauthorrefmark{1}, 
Belkacem~Mouhouche\IEEEauthorrefmark{2},
Pei~Xiao\IEEEauthorrefmark{1} and Rahim~Tafazolli\IEEEauthorrefmark{1}}
	\IEEEauthorblockA{\IEEEauthorrefmark{1}Institute for Communication Systems, University of Surrey, United Kingdom}
	\IEEEauthorblockA{\IEEEauthorrefmark{2}Samsung Electronics R{\&D} UK, United Kingdom}
	Email:\{hongzhi.chen, d.mi, p.xiao, r.tafazolli\}@surrey.ac.uk, b.mouhouche@samsung.com} 



%


\maketitle
\begin{abstract}
This paper provides the mobility and coverage evaluation of New Radio (NR) Physical Downlink Control Channel (PDCCH) for Point-to-Multipoint (PTM) use cases, e.g., eMBMS (evolved Multimedia Broadcast Multicast Services). The evaluation methodology is based on analyses and link level simulations where the channel model includes AWGN, TDL-A, TDL-C as well as a modified 0dB echo to model different PTM scenarios. The final version of this work aims to provide insightful guidelines on the delay/echo tolerance of the NR PDCCH in terms of mobility and coverage. 
In this paper, it is observed that under eMBMS scenario, i.e. SFN channel, due to the time domain granularity of pilots distributed inside the PDCCH region, the system can support very high user movement speed/Doppler with an relatively low requirement on the transmit Signal/Carrier-to-Noise Ratio (SNR/CNR). On the other hand however, the system falls short on its coverage due to the low frequency domain granularity of pilots that effects the channel estimation accuracy.
\end{abstract}


%
\IEEEpeerreviewmaketitle

{\scriptsize \footnote{We would like to acknowledge the support of the University of Surrey 5GIC (www.surrey.ac.uk/5gic) members for this work. This work was also supported in part by the European Commission under the 5GPPP project 5G-Xcast (H2020-ICT-2016-2 call, grant number 761498). The views expressed in this contribution are those of the authors and do not necessarily represent the project.}}


\section{Introduction}\label{sec:intro}
In the 3GPP (3rd Generation Partnership Project) standardization of 5G (5th Generation) New Radio (NR), the design of control channels has been through significant changes compared with that in LTE (Long Term Evolution) Advanced Pro, in order to better handle NR data streams. Most changes of the main type of the physical control channels, i.e., Physical Downlink Control Channel (PDCCH), have been captured in our prior work \cite{GC18WS, TBC, SAM}. In this work, leveraging on our current project \cite{D31, D32}, we aim at evaluating the performance of the PDCCH in Point-to-Multipoint (PTM) scenarios, as a potential 5G broadcast and multicast solution. This work utilizes as reference the Key Performance Indicators (KPI) and evaluation methodology defined by the ITU-R (International Communications Union - Recommendation) for the IMT-2020 (International Mobile Telecommunication) evaluation process \cite{ITU_R_Guidelines}, in which it only takes into account the Point-to-Point (PTP) scenarios. 
By focusing on the multimedia broadcast and multicast services, the performance of the New Radio specified Physical Downlink Shared Channel (PDSCH) was studied in \cite{NRDATA}. As well as the Block Error Rate (BLER) of the Downlink Control Information (DCI) for the PDCCH in the PTP scenario was studied in \cite{GC18WS} which assumes static receiver and AWGN channel. 
Methods/designs that aiming to enhance the performance/reliability of the PDCCH under PTP environment were also studied in the literature. For example, a potential power-based optimization for the control information transmission based on LTE has been discussed in \cite{powerdci} and can be extended into the NR scenario. In \cite{one5g1, one5g2}, the first one studies the BLER of the PDCCH with two
types of transmit diversity schemes and different diversity orders. The second one demonstrates that SFBC based transmit diversity outperforms per-RE precoder cycling scheme. The second one presents the
exponential effective SNR mapping (EESM) results of the configurable distributed PDCCH, and studies the
the trade-off between achieved frequency diversity and channel estimation gain with different
adjusting resource bundling level.

In this work, we present a comprehensive performance evaluation of both mobility and coverage for point-to-multipoint (PTM) systems in order to exam the delay/coverage tolerance for the PDCCH transmission on the current control channel configuration. 
The analysis and simulation follow the physical layer chain defined by 3GPP in \cite{TR36212} and \cite{TR38211}. And the results are compared with the mobility and coverage requirement that are defined in \cite{ITU_R_Guidelines} and \cite{D21}. The simulation results show that based on the current pilot distribution in the PDCCH area, control information can be reliably transmitted under a very wide range of user mobility under a 5G channel i.e., TDL-A. Both ideal and real channel estimation cases have been covered in the coverage evaluation. It shows that finding a suitable interpolation method is very important in order to reconstruct the DCI due to the lack of pilots compare to the channel delay spread.   

This paper will be organized as follows. First, it describes the NR PDCCH framing, Transmitter side processing chain and the modified channel modeling to model the PTM scenarios. Next, it presents the analysis as well as link level simulation results, focusing on the delay/echo tolerance of the NR PDCCH in terms of mobility and coverage. Finally, it summarizes the key findings of the investigations carried out and discusses the potential improvements towards the development of the next generation PTM technologies.
\section{Point-to-Multipoint PDCCH Initialization}\label{sec:sys}
\subsection{PDCCH Frame Structure}
\begin{figure}[t]
	\centering
	\includegraphics[width=0.48\textwidth]{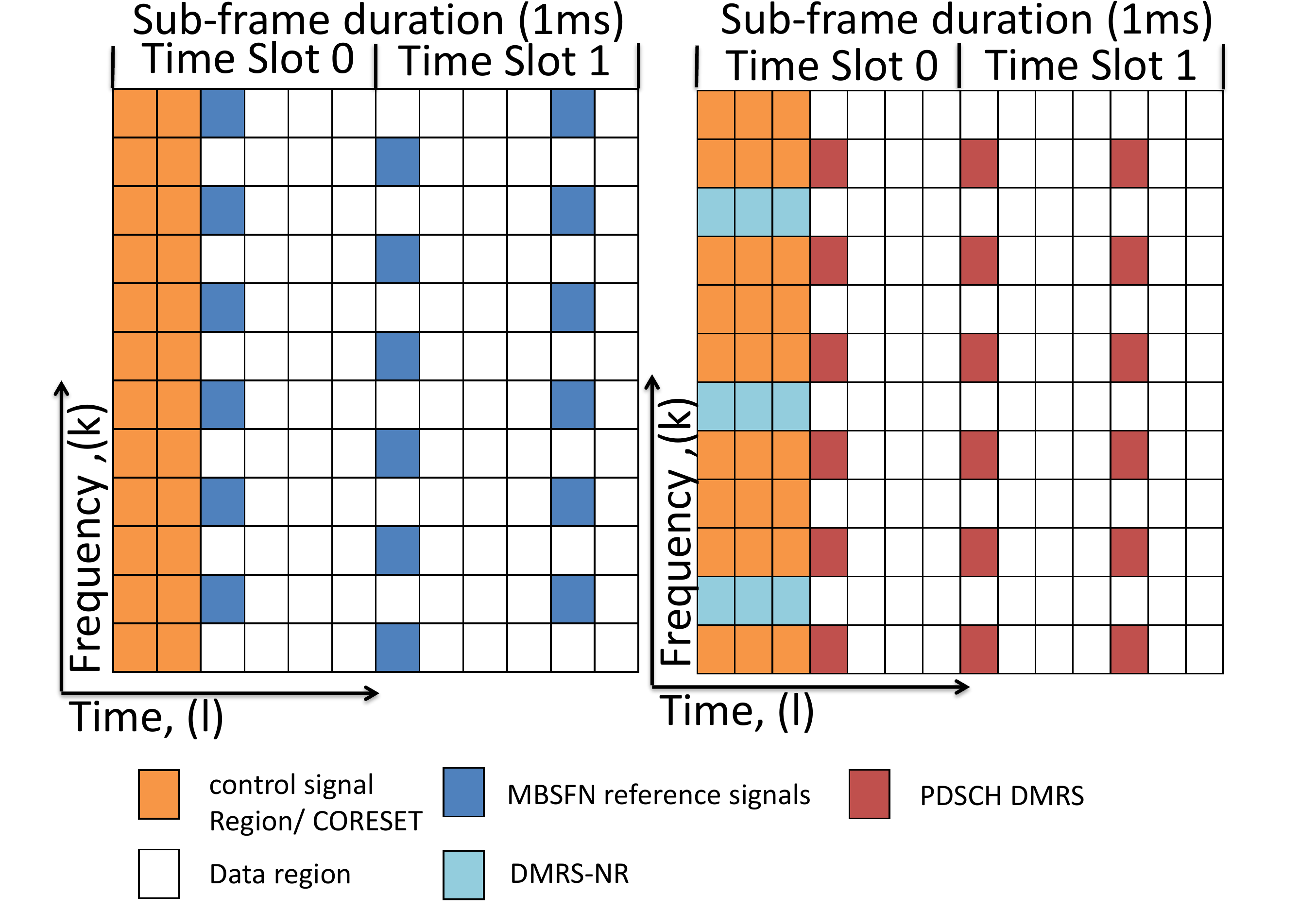}
	\caption{Resource Block structure in LTE-MBSFN (left) and NR (right), with 15 kHz carrier spacing.}
	\label{Fig_Framestructure}
\end{figure} 
The control information specifies the data scheduling and allocation for each user equipment (UE) by means of Downlink Control Information (DCI). In NR, the DCI is mapped to Control Resource Set (CORESET), in which its content can be distributed at most over three consecutive Orthogonal Frequency Division Multiplexing (OFDM) symbols, depending on high layer parameters. CORESET also includes Demodulation Reference Signal (DMRS) for the correct demodulation of the PDCCH. The right-side figure of Fig.~\ref{Fig_Framestructure} illustrates one possible DMRS/pilot pattern.
\subsection{Transmitter Side Block Diagram}
\begin{figure}[h]
	\centering
	\includegraphics[width=0.49\textwidth]{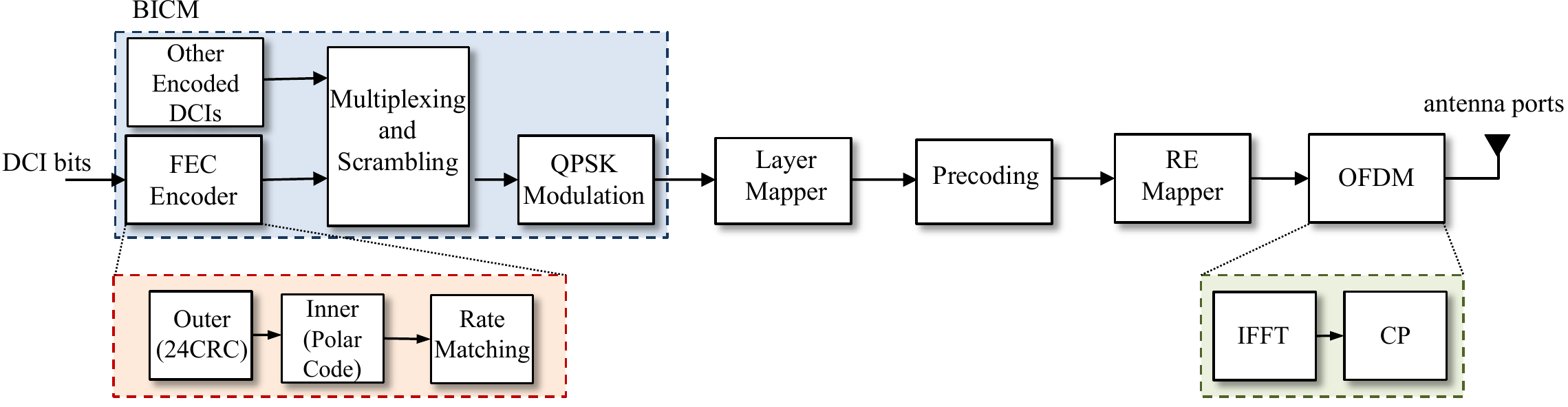}
	\caption{NR physical layer PDCCH transmit block diagram.}
	\label{blockdiagram}
\end{figure}
The PDCCH processing chain/transmit block diagram of the NR PDCCH processing chain is shown in Fig.\ref{blockdiagram}. Therein, the generator polynomial $g_\text{CRC24C}(D)$ is used for the Cyclic Redundancy Check Attachment. Polar codes are used for channel coding, where the details of polar encoding in NR can be found in \cite{TR38211}. It is worth mention that, the length of the encoded bits is $N=2^n$, where $n$ is a integer between 5 and 9, and the maximum length of the encoded bits is fixed at $2^9=512$. Rate matching consists of sub-block interleaving, bit collection, and bit interleaving. Same as in 4G LTE the modulation scheme that is operated on the PDCCH is QPSK.
\subsection{0dB Echo Channel Modelling}
\begin{table}[t]
	\centering
	\caption{Power Delay Profile for Modified Echo Channel}
	\renewcommand{\arraystretch}{1.2}
	\begin{tabular}{||c|c|c||} 
		\hline
		Path & Amplitude(dB) & Delay($\mu s$) \\
		\hline\hline
		1 & 0 & 0\\ 
	    2 & 0 & $\alpha$*$T_{cp}$ \\
		\hline
	\end{tabular}
	\label{T1}
\end{table}
Table \ref{T1} includes the power delay profile (PDP) of the 0dB echo channel, where tuning the value of variable $\alpha\geq0$ is equivalent to changing the coverage of Single Frequency Network (SFN) area, such that we can evaluate the performance for different PTM scenarios, and consequently the delay/echo tolerance of the NR PDCCH in terms of mobility and coverage.

\section{Analysis and Link-Level Simulation Evaluation}\label{sec:sim}
This section studies the speed tolerance of the New Radio PDCCH channel with practical channel estimation algorithms as well as the corresponding link level simulation results to valid the analysis.
\subsection{Doppler Effect}
\subsubsection{Theoretical Doppler Limit}
\begin{figure}[t]
	\centering
	\includegraphics[width=0.47\textwidth]{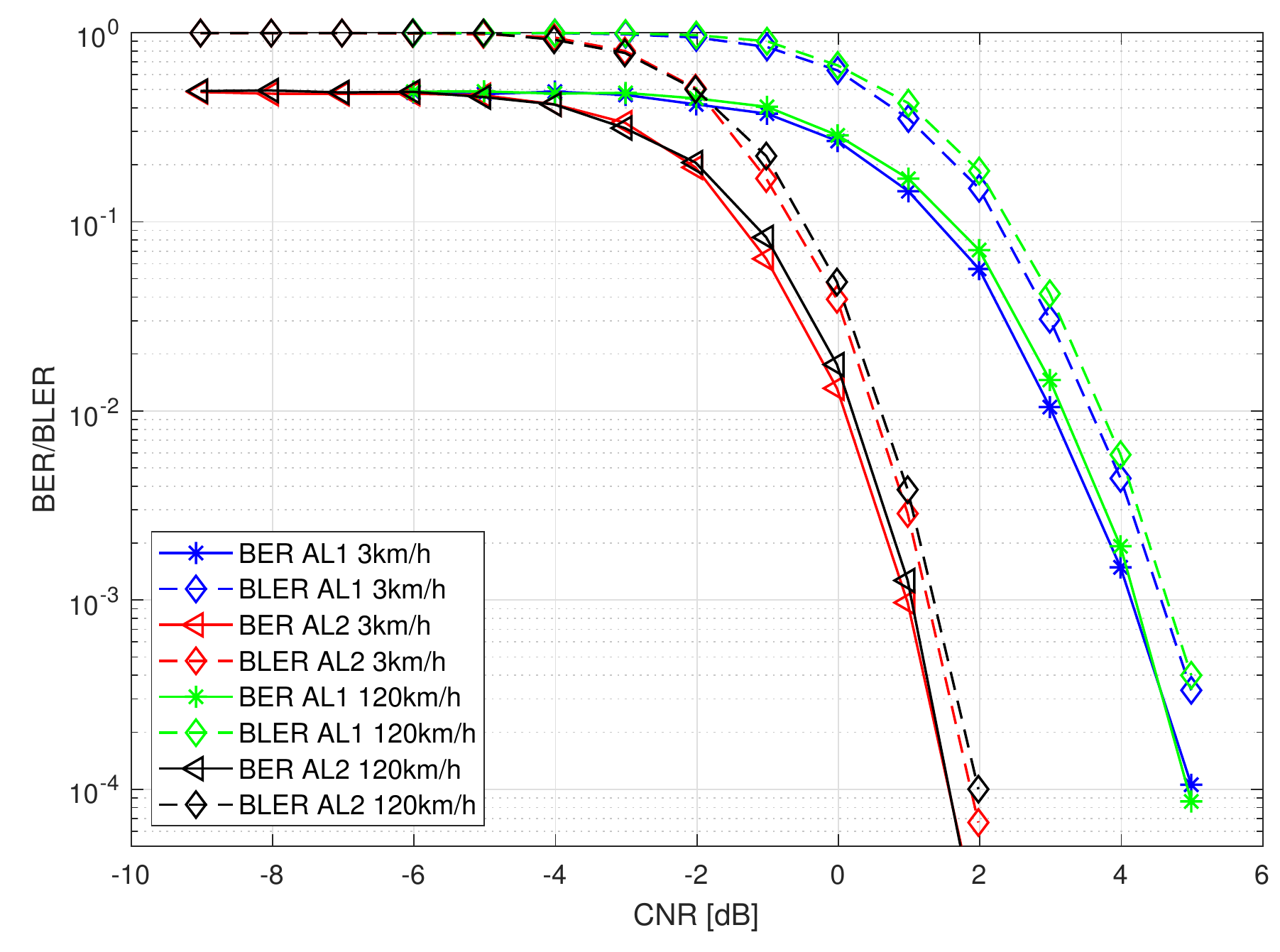}
	\caption{BLER/BER vs. CNR for Aggregation Level 1\&2 with different user speed}
	\label{MobilityBLER}
\end{figure} 
\begin{figure}[t]
	\centering
	\includegraphics[width=0.47\textwidth]{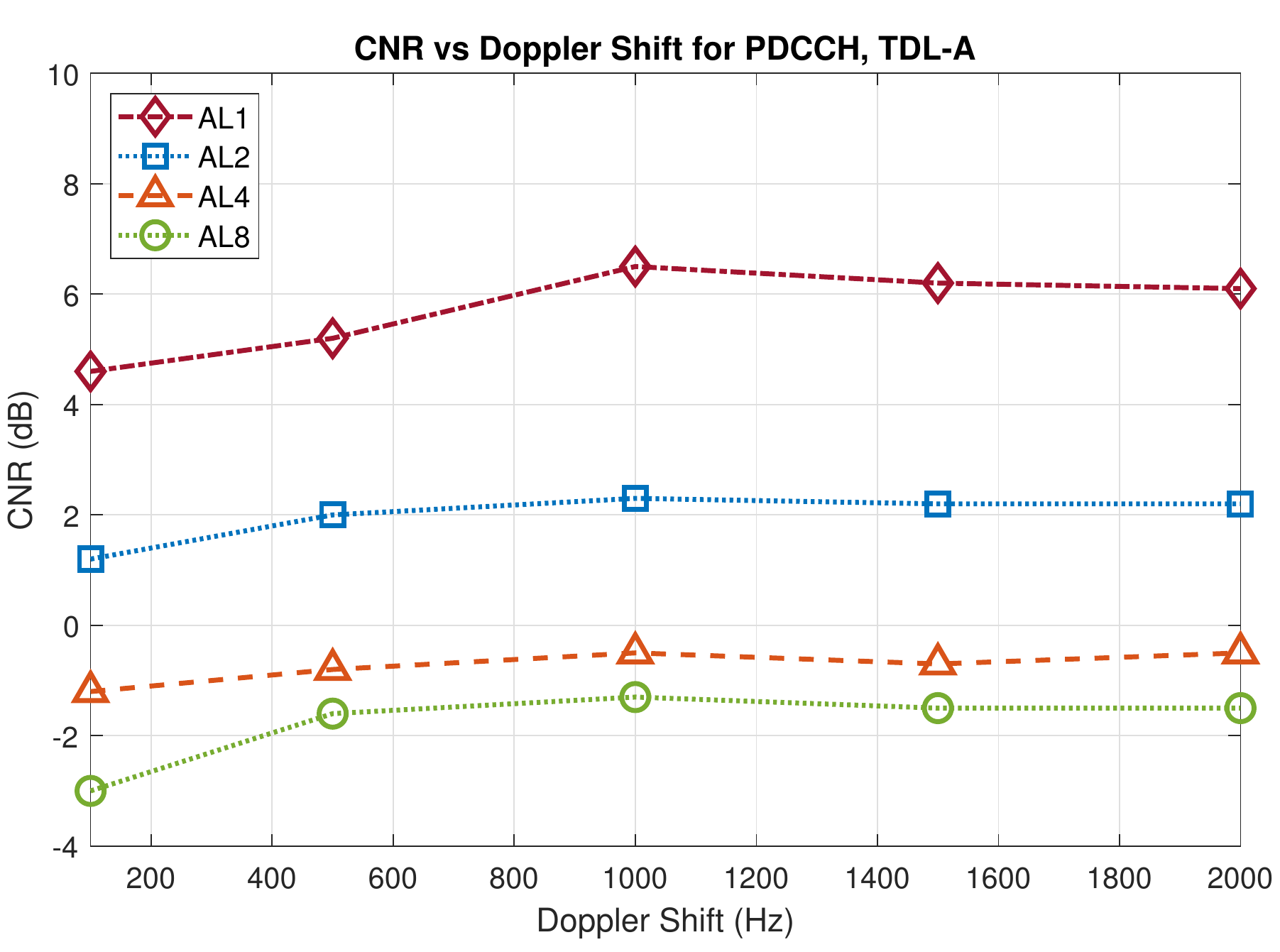}
	\caption{PDCCH CNR vs. Doppler Shift (user speed) for 5G New Radio in TDL-A mobile channel with real channel estimation.}
	\label{Mobility}
\end{figure} 

In order to perform an effective channel estimation, a two-dimensional (i.e., frequency and time) sampling should satisfy:
\begin{itemize}
	\item In the frequency domain, the sampling rate must be faster than or equal to the maximum delay spread of the channel;
	\item In the time domain, the sampling rate must be greater than or equal to the maximum Doppler spread of the channel.
\end{itemize}
Based on the above-mentioned condition, the maximum distance between two PDCCH DMRS symbols in the time domain, $n_{max}$, is given by
\begin{equation}
n_{max}\leq \frac{1}{2*(T_U+T_{cp})*d_{max}},
\end{equation}
where $T_U$ is the useful symbol duration, $T_{cp}$ is the cyclic prefix (CP) length in time, and $d_{max}$ represents the maximum Doppler spread of the channel. Due to the fact that the PDCCH DMRS symbols cover all resource elements (REs) in time domain on selected subcarriers, thus the number of PDCCH DMRS symbols is sufficient enough to capture the time variation of the channel, i.e., to be able to potentially support a wide range of user speeds, in the sense that
\begin{equation}
f_p = \frac{1}{2D_y(T_U+T_{cp})} \text{Hz},
\end{equation} 
where $f_p$ is the maximum frequency shift supported, $D_y$ is the length of the PDCCH DMRS in OFDM symbols, and $D_y = 1$ in this case. Hence, 
\begin{equation}
	f_p = \frac{1}{2*1*10^{-6}*(66.67+5.2)} = 6963.6 \text{Hz},
\end{equation}
which corresponds to a user speed of 10743 km/h at 700 MHz and 1880 km/h at 4 GHz frequency. Note that the higher numerologies will increase the supported UE speed, since lower $T_U+T_{cp}$ is obtained.\\
\subsubsection{Link-level Evaluation}

To evaluate the performance of PDCCH in the mobile environments, link level simulations are performed to, first, verify if any error floor occurs in the Bit Interleaving, Coding and Modulation (BICM) performance when the estimated channel response applies and different user speeds are considered. If not, then the required Carrier-to-Noise Ratio (CNR) to achieve Block Error Rate (BLER) $<$ 0.1\% against the Doppler shift or user speeds can be evaluated. The two-dimensional pilot-based estimation with the linear interpolation is used. Different aggregation levels have been considered. 

In Fig.\ref{MobilityBLER}, the BLER/BER vs. CNR results for aggregation level 1 and 2 with user speed of 3km/h and 120 km/h are presented. From Fig.\ref{MobilityBLER} we can see that the required CNR to achieve 1e-3 BLER just slightly right shifted for the 120km/h (i.e., about 4.5dB CNR) case compare to the 3km/h case (4.4dB CNR) and no error floor occurs, which verify the above-discussed doppler effect based on the sampling rate requirement.  
Therefore, the required CNR vs. a wide range of Doppler Shift (equivalent to a wide range of user movement speed) are shown in Fig. \ref{Mobility}. From Fig. \ref{Mobility}, we can see that the PDCCH channel can handle all required user speeds with a slightly increased CNR requirement for all aggregation levels, as in \cite{D21} for all the considered frequency bands. Also, the higher the aggregation level, the lower the required CNR, due to the better coding rate used (half with the next level). 

\subsection{Coverage}
In this sub-section, the performance of the NR-PDCCH channel with different coverage settings has been evaluated. 
\subsubsection{With Perfect Channel Estimation}
\begin{figure}[t]
	\centering
	\includegraphics[width=0.48\textwidth]{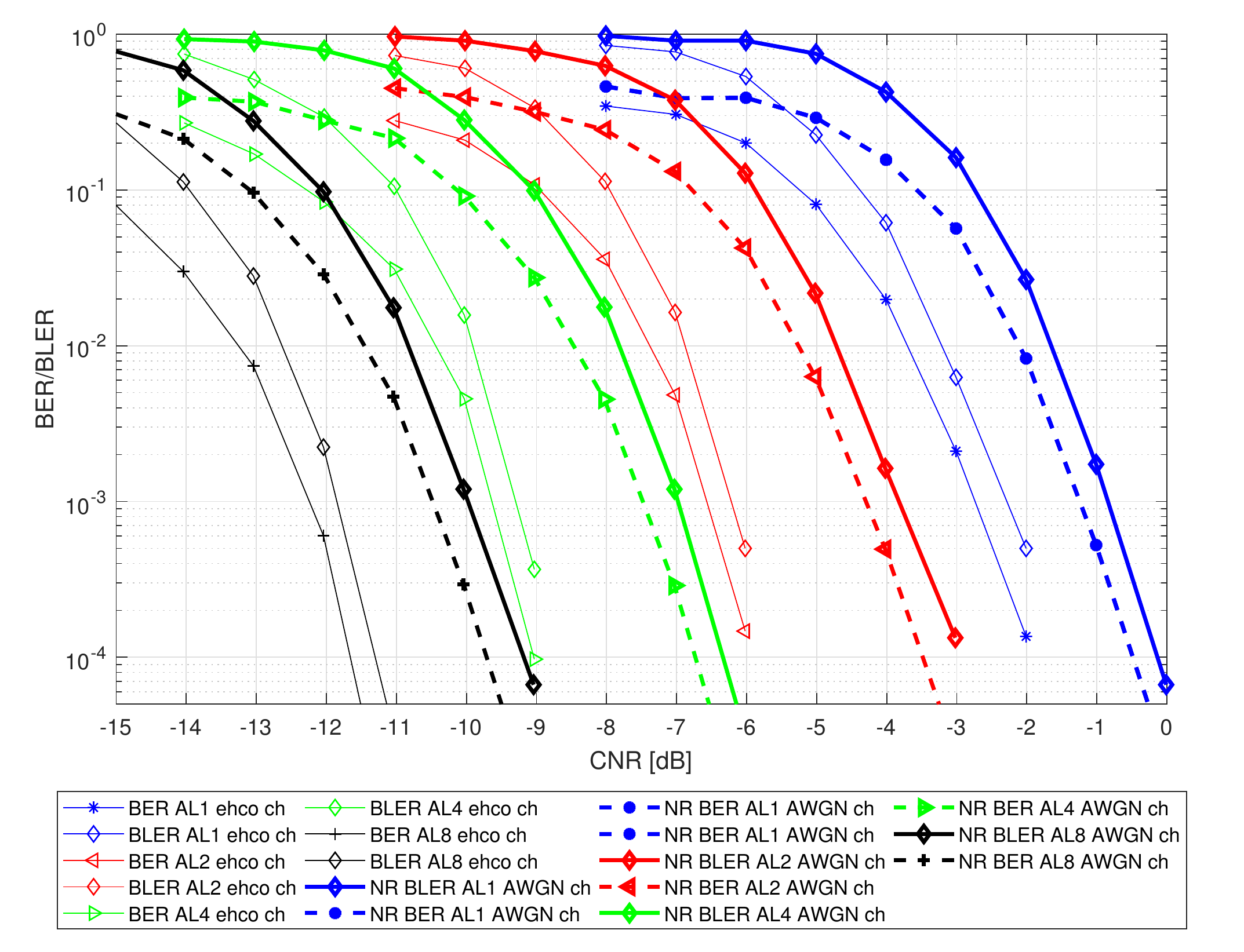}
	\caption{BER/BLER vs. CNR for AWGN and 0dB echo channel with perfect channel estimation}
	\label{NOCE}
\end{figure}

First, we take a look at the BLER/BER results for both AWGN and 0dB each channel ($\alpha=0.3$ and this represents the SFN receiver) with perfect channel knowledge, and the results are shown in Fig. \ref{NOCE}. From Fig. \ref{NOCE}, we conclude that a higher aggregation level generally gives more protection level to the codewords, which is reflected on the required CNR for both AWGN and 0dB echo channel, but with less spectrum efficiency. Breaking down to each aggregation level, to achieve a BLER level of 1e-3, for AWGN channel, it requires CNR values of -0.9, -3.8, -7, -10dB for aggregation level 1, 2, 4, 8, respectively. While for 0dB each channel, the corresponding CNR values are -2.3, -6.2, -9.3, -11.9dB, respectively. However, ideal channel estimation cannot reflect the pilot granularity when different channel delay is implied. So in the next subsection, we will introduce the results with real channel estimation.
\subsubsection{With real Channel Estimation}

 In theory, with a specific numerology $\mu$ and therefore a fixed useful symbol duration $T_U$, the maximum channel delay spread is highly dependent on the DMRS pattern, or to be specific, the pilot granularity in the frequency domain $D_x$. The maximum distance between two frequency domain PDCCH DMRS symbols, $m_{max}$, is given by:
\begin{equation}
	m_{max}\leq \frac{T_U}{\tau_{max}}
\end{equation}
where $\tau_{max}$ represents the maximum delay spread of the channel. Take the 15kHz subcarrier spacing as an example, also as shown in the frame structure in Fig.\ref{Fig_Framestructure}, $D_x=4$, one can derive $T_U=\frac{1}{\Delta_f} =66.7\mu s$. Therefore, the maximum channel delay spread that can be tolerated would be $\tau_{max}\leq \frac{T_U}{D_x}\approx16.67\mu s$, greater than the one from the channel such as TDL-A and TDL-C. Thus, similar to the NR-PDSCH, the 0dB echo channel has been modified and extended to include different delays. 
Regarding the evaluation, in terms of interpolations after channel estimation on the pilot, there is no need to perform interpolations in time domain due to the corresponding pilot distribution in the PDCCH region. For the frequency domain, the DFT-based interpolation is applied, which is a widely-used channel estimation method and shows promise in \cite{DFT}, compared with the estimator with the other types of interpolations such as the linear interpolation. Two set of simulations are considered coverage capability:
\begin{itemize}
	\item Normal CP length with 7\% Fast Fourier Transform (FFT) size (only for $\mu = 0$).
	\item Extended CP with 25\% of FFT size. 
\end{itemize}
\begin{figure}[t]
	\centering
	\includegraphics[width=0.49\textwidth]{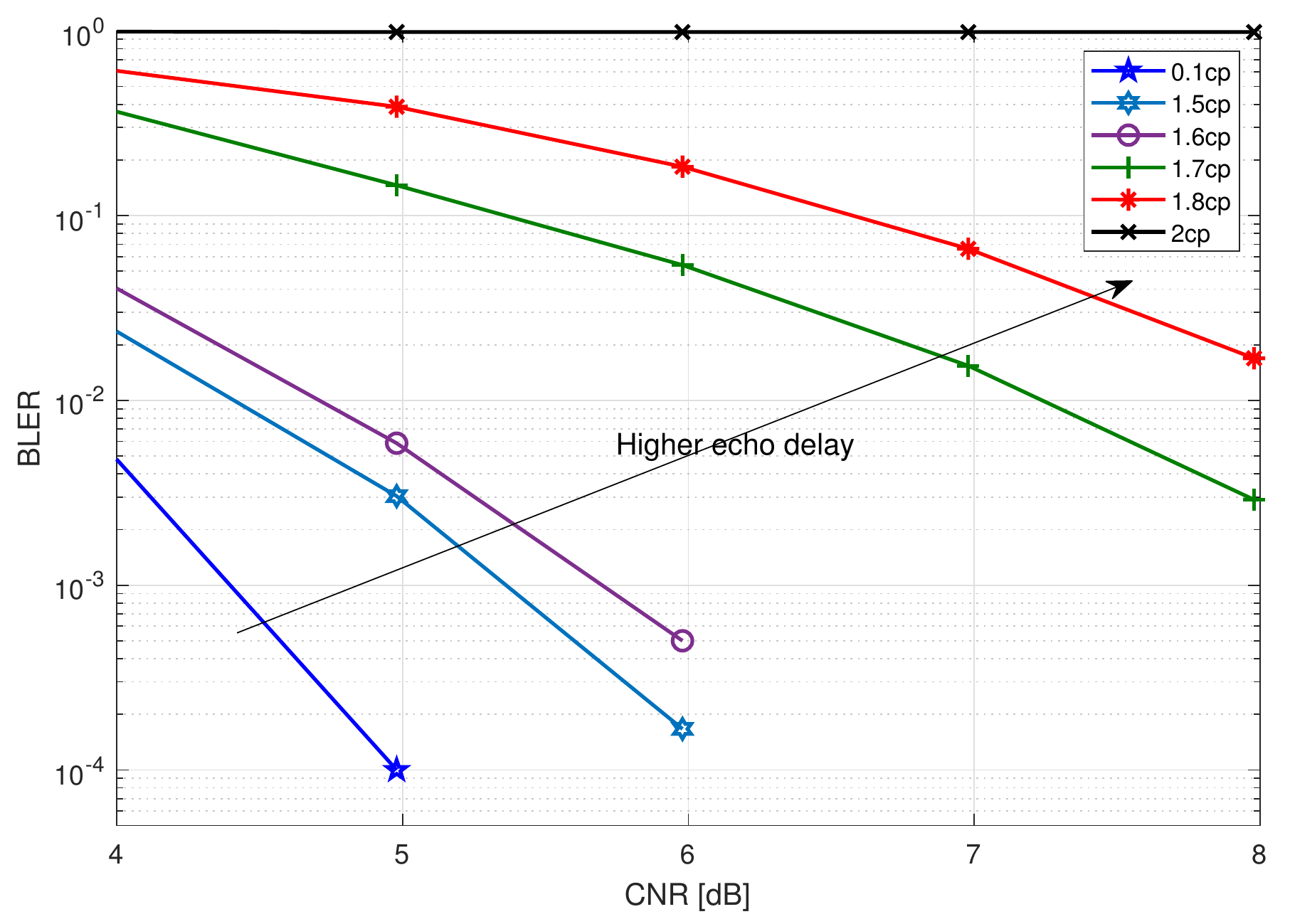}
	\caption{BLER vs. CNR with different echo delay (PDCCH, AL=2).}
	\label{BLERCNRecho}
\end{figure} 
\begin{figure}[t]
	\centering
	\includegraphics[width=0.49\textwidth]{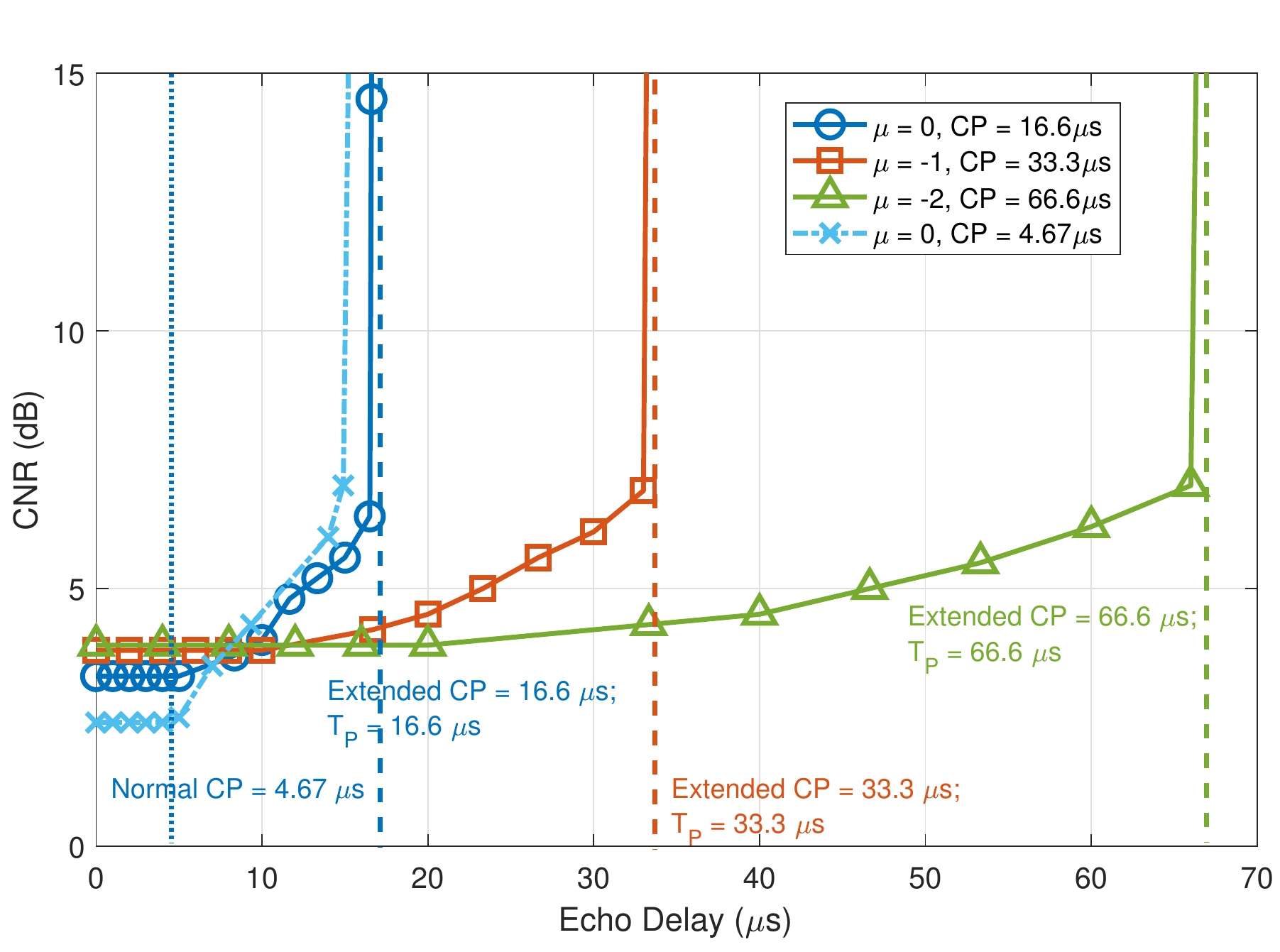}
	\caption{Required CNR vs. corresponding echo delay (PDCCH, AL=2).}
	\label{coverage}
\end{figure} 
The pilot granularity remains the same in both scenarios, and the aggregation level 2 is considered.
First, we investigate the BLER vs. CNR for aggregation level 2 with different echo delay i.e. $\alpha = 0.1, 1.5, 1.6, 1.7, 1.8, 2$ and the frame structure for numerology $\mu = 0$ is used. From Fig. \ref{BLERCNRecho} we can see when the $\alpha$ go beyond 1.6, the required CNR dramatically increased (from 5.6dB to more than 8dB in order to achieve 1e-3 BLER level) and the link becomes unreliable when $\aleph = 2$.  Same simulations have been executed for extended CP with different numerologies. In Fig.\ref{coverage} we present the corresponding results which show the trend of required CNR (to achieve 1e-3 BLER) following the increasing of echo channel delay. 

With the numerology $\mu = 0$, both normal and extended CPs can work properly with the time delay equivalent to the normal CP duration. Then the larger the time delay, the higher the required CNR for normal CP, due to the increased inter-symbol interference (ISI). However, the required CNR for the extended CP is increased even within the CP duration. Regarding the maximum delay spread to be tolerated, both normal CP and extended CP can support up to around 16$\mu s$, aligned with the theoretical limit as calculated previously. Focusing on the numerology $\mu = 0$, the usage of extended CP does not help to increase the PTM coverage and the required CNR to achieve the same BLER is increased with extended CP. Two insights can be highlighted:
\begin{itemize}
	\item CNR loss due to increased CP length
	\item Low pilot granularity in frequency domain that affect the channel estimation accuracy
\end{itemize}
\begin{figure}[t]
	\centering
	\includegraphics[width=0.49\textwidth]{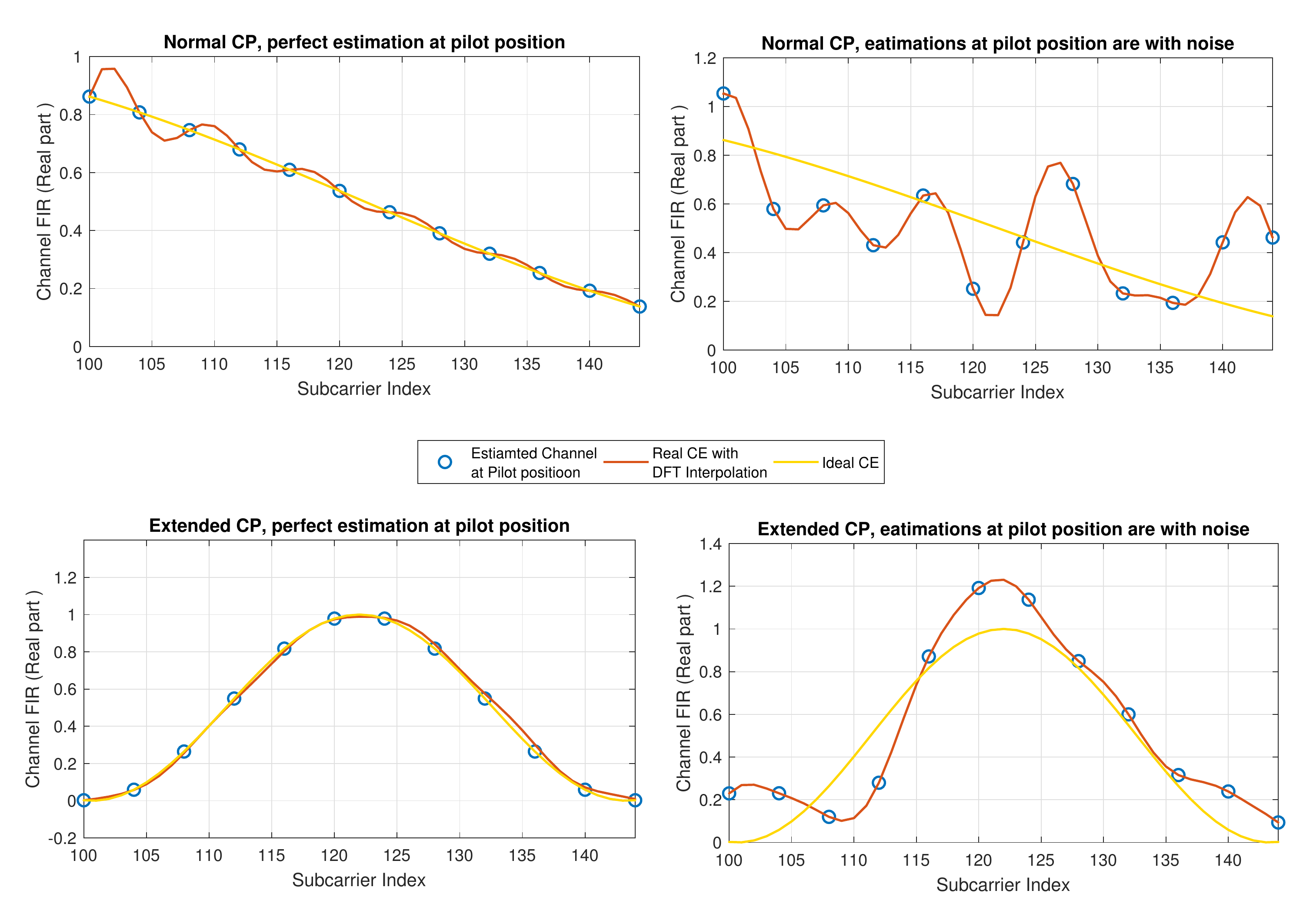}
	\caption{Normal \& extended CP with 0.1*$T_{cp}$ as second path delay}
	\label{0.1cp}
\end{figure} 
For the first point, as we know, the insertion of CP disperses the transmitter energy (the amount of consumed power depends on how large the CP length), where the signal-to-noise (SNR) lost due to the CP introduction indicates the loss of transmission energy. The loss factor is given as:
\begin{equation}
	\text{CNR}_{loss} = -10\log_{10}(1-\frac{T_{cp}}{T_{U}+T_{cp}}),
\end{equation}
which is equivalent to the increase of noise variance (if fixed symbol energy.), thus yields an increased channel estimation error at the pilot position. As we can see the comparison in Figure 29, with a relatively short 2nd path delay, both normal and extended CP scenarios can still reconstruct the channel within an acceptable offset range. However, in Figure 30, with higher 2nd path delay (but still inside the CP duration), with normal CP, the channel can be reconstructed, which is not the case for the extended CP. Combining with the second aspect, the setting with the extended CP results in that the channel is almost unable to be reconstructed.
Especially for the second point, when comparing the two figures at the left side of Fig.~\ref{0.1cp}, with normal CP, the interpolation nearly perfectly captures the channel variant. However, with extended CP, even with perfect channel estimation at the pilot position, the interpolation cannot reconstruct the channel due to the low pilot granularity, and the gap is further amplified by the equivalently increased noise variance, as shown in the right two figures in Fig.~\ref{0.6cp}. To elaborate more on the relationship between the pilot granularity in the frequency domain and the supported maximum delay spread of the channel, we consider the requirement of the DFT-based channel estimation/interpolation. As discussed in \cite{DFT}, it requires the number of pilots to be much greater than the channel delay spread (counted as the number of the channel delay taps in the time domain). The frequency domain pilot granularity in the current PDCCH DMRS pattern cannot meet this requirement in certain scenarios, which can cause the imperfection of the DFT-based channel estimation, thus the performance degradation of the system.
\begin{figure}[t]
	\centering
	\includegraphics[width=0.49\textwidth]{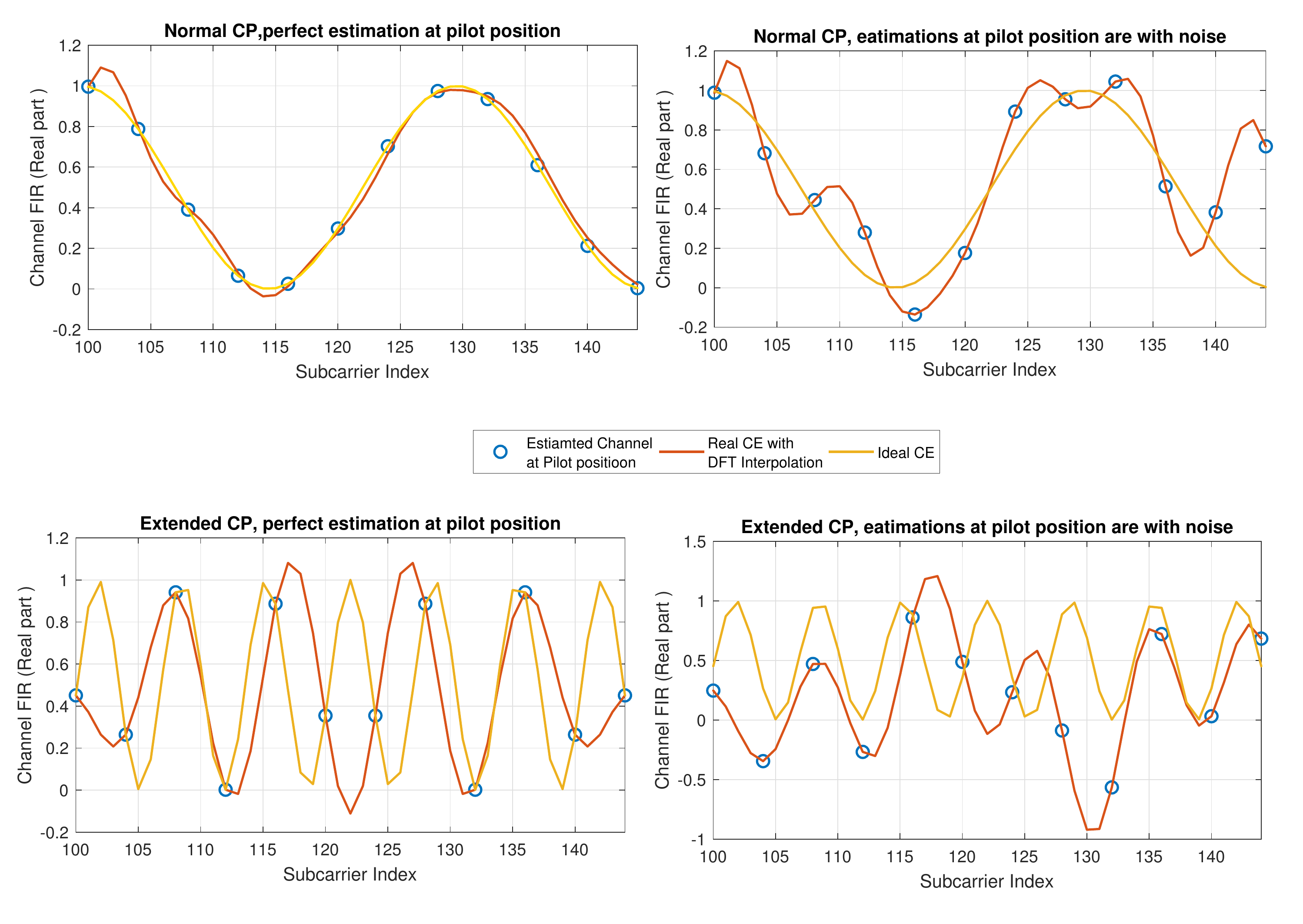}
	\caption{Normal \& extended CP with 0.6*$T_{cp}$ as second path delay}
	\label{0.6cp}
\end{figure} 
\section{Possible Enhancement}\label{sec:possenhance}
\begin{figure}[t]
	\centering
	\includegraphics[width=0.49\textwidth]{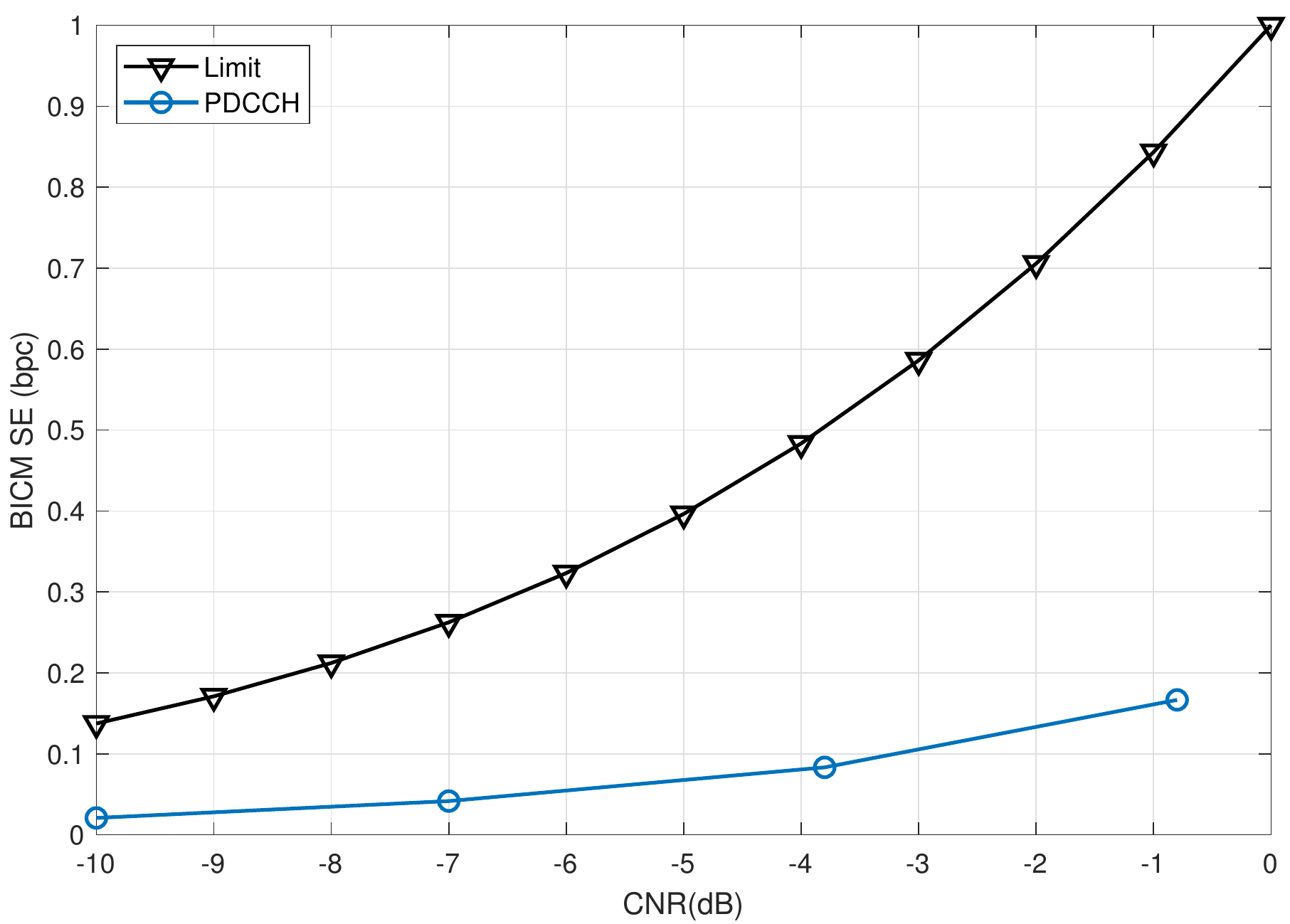}
	\caption{BICM Spectral Efficiency vs. CNR for AWGN channel, Capacity vs. PDCCH}
	\label{PDCCHCE}
\end{figure}

One solution to enhance the coverage capability is to use the negative numerology as proposed previously. Another possible solution is to design a new PDCCH DMRS pattern, especially for the extended CP. Recall the relationship between the distance between two frequency domain PDCCH DMRS symbols and the supporting delay spread of the channel, one can design the control channel frame or the DMRS pattern by decreasing $D_x$. This, however, can result in the significantly decreased number of available REs for the control channel payload transmission, considering the limited number of available radio resources and very low spectral efficiency in the control channel. Besides, only QPSK can be used, which capped the peak SE to 2 bits per channel user. Shown in Fig. \ref{PDCCHCE} is the spectral efficiency of the current standard, as we can see even in AWGN channel, the spectral efficiency is already far away from capacity. 

\section{Conclusion}\label{sec:conclusion}
In this paper, the Doppler effect limitation is analyzed based on the PDCCH pilot distribution as well as the coverage which is covered by using the 0dB echo channel to represent the SFN receiver. 
The mobility and coverage simulation of PDCCH performance has been evaluated for AWGN, as well as for TDL-A channel models. 
The discussions and simulation results obtained in this paper shows that based on the current pilot distribution inside the PDCCH area, it can support a very wide range of user mobility (beyond the requirement for a PTM scenario) under TDL-A channel. Regarding coverage, both ideal and real channel estimation cases have been covered. It shows that finding a suitable interpolation method is very important in order to reconstruct the DCI due to the lack of pilots compare to the channel delay spread. 
Besides, as a potential future work, the imperfect PDCCH recovery at the receiver may lead to a performance loss for the PDSCH performance, so these two channels can be combined together to get a more completed link level simulation. A comparison with the signaling of the state-of-the-art DTT standard ATSC 3.0 would be of interest too.




%
\bibliographystyle{IEEEtran}
\bibliography{reference}

\end{document}